\documentclass[]{cupconf}
\usepackage{graphicx}

\title[The metallicity distribution of stars in ellipticals.]
      {The metallicity distribution of the stars in elliptical galaxies}
\author[A.Pipino \& F.Matteucci]{Antonio Pipino, Francesca Matteucci}
\affiliation{Dipartimento di Astronomia, Universit\'a di Trieste, Italy}

\begin{document}
\maketitle

\begin{abstract}

Elliptical galaxies probably host the most metal rich stellar populations in the Universe.
The processes leading to both the formation and the evolution of such stars
are discussed by means of a new multi-zone photo-chemical evolution model,
taking into account detailed nucleosynthetic yields, feedback from supernovae, Pop III stars and an initial infall
episode. Moreover, the radial variations in the metallicity distribution
of these stars are investigated by means of G-dwarf-like diagrams.

By comparing model predictions with observations, we derive a
picture of galaxy formation in which the higher is the mass of the
galaxy, the shorter are the infall and the star formation
timescales. Therefore, the stellar component of the most massive and luminous galaxies
might attain a metallicity $Z \ge Z_{\odot}$ in only 0.5 Gyr.

Each galaxy is created
outside-in, i.e. the outermost regions accrete gas, form stars and
develop a galactic wind very quickly, compared to the central core in
which the star formation can last up to $\sim 1.3$ Gyr.  This
finding will be discussed at the light of recent observations of the
galaxy NGC 4697 which clearly show a strong radial gradient in
the mean stellar [$<Mg/Fe>$] ratio.

\end{abstract}

\firstsection
\section{Introduction}
Metallicity gradients are characteristic of the stellar populations
inside elliptical galaxies. Evidences come from the increase of
line-strength indices (e.g. Carollo et al., 1993; Davies et al., 1993;
Trager et al., 2000) and the reddening of the colours (e.g. Peletier
et al. 1990) towards the centre of the galaxies.  The study of such
gradients provide insights into the mechanism of galaxy
formation, particularly on the duration of the chemical enrichment
process at each radius.  Metallicity indices, in fact, contain
information on the chemical composition and the
age of the single stellar populations (SSP) inhabiting a given
galactic zone.  
Pipino \& Matteucci (2004, PM04)
showed that a galaxy formation process in which the most massive
objects form faster and more efficiently than the less massive ones
can explain the photo-chemical properties of ellipticals, in 
particular the increase of [Mg/Fe] ratio in stars with galactic mass 
(see Matteucci, this conference, and references therein).
PM04 suggested that a single galaxy should form outside-in, namely the
outermost regions form earlier and faster with respect to the central
parts.  A natural consequence of this
model and of the time-delay between the production of Fe and that of Mg
is that the mean [Mg/Fe] abundance ratio in the stars should
increase with radius. 
Pipino et al. (2006, PMC06) compared PM04 best model results with the very recent
observations for the galaxy NGC 4697 (Mendez et al. 2005), and found
them in excellent agreement.

\section{The model}
The chemical code adopted here is described in full detail 
in PM04 and PMC06, where we address the reader for more details. 
This model is characterized by:
Salpeter (1955) IMF, Thielemann et al. (1996) yields for massive stars,
Nomoto et al. (1997) yields for type Ia SNe and 
van den Hoek \& Groenewegen (1997) yields for low-
and intermediate-mass stars (the case with $\eta_{AGB}$ varying with metallicity). 
Here we present our analysis of a $\sim 10^{11}M_{\odot}$ galaxy (PM04 model IIb), considered
representative of a typical elliptical, unless otherwise stated.

The model assumes that the galaxy assembles by merging of gaseous
lumps (infall) on a short timescale and suffers a strong star burst
which injects into the interstellar medium a large amount of
energy able to trigger a galactic wind, occurring at different
times at different radii.  After the development of the
wind, the star formation is assumed to stop and the galaxy evolves
passively with continuous mass loss.

\section{Results and discussion}

\begin{figure*}
\includegraphics[width=2.6in,height=2.4in]{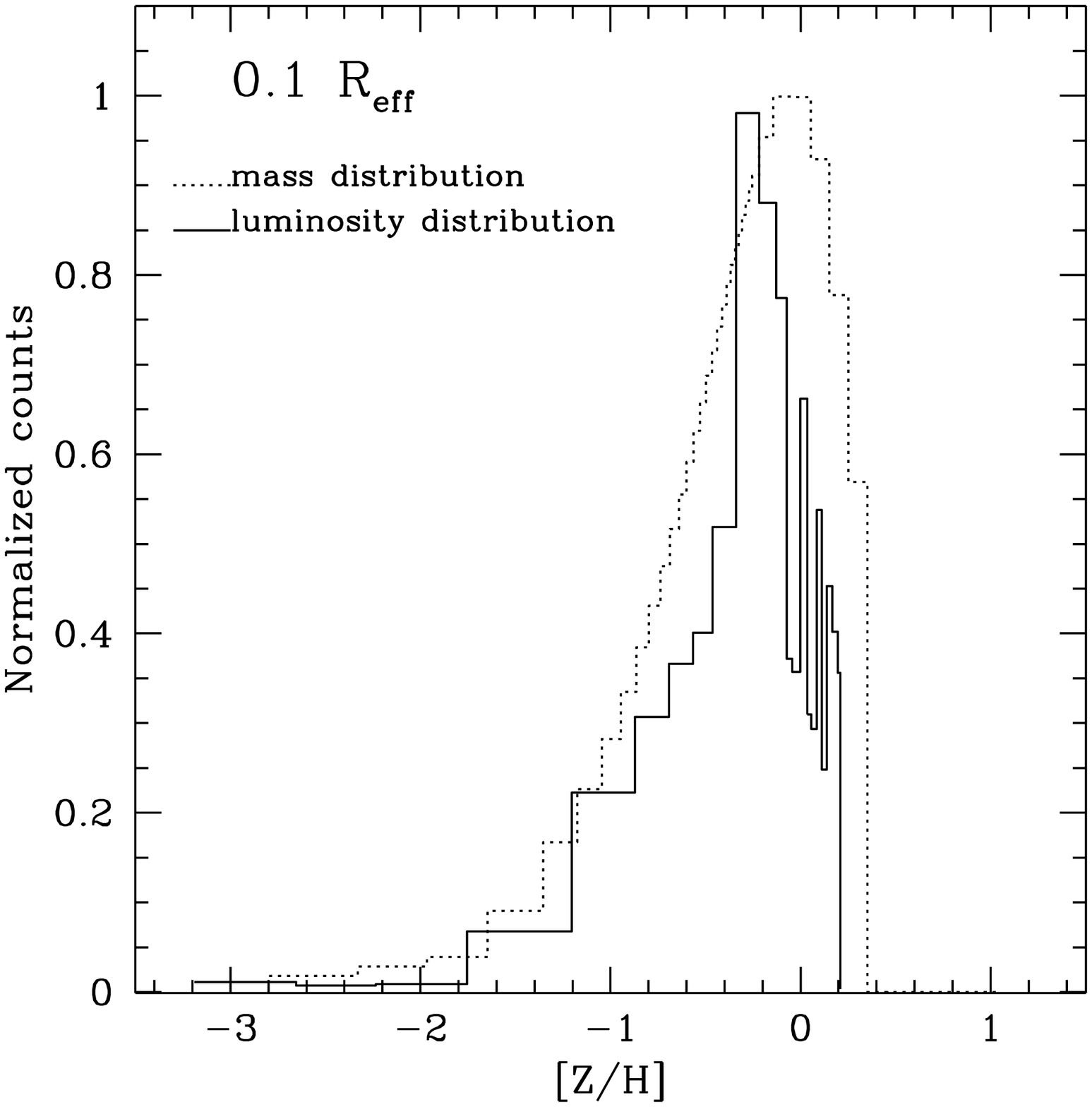}
\includegraphics[width=2.6in,height=2.4in]{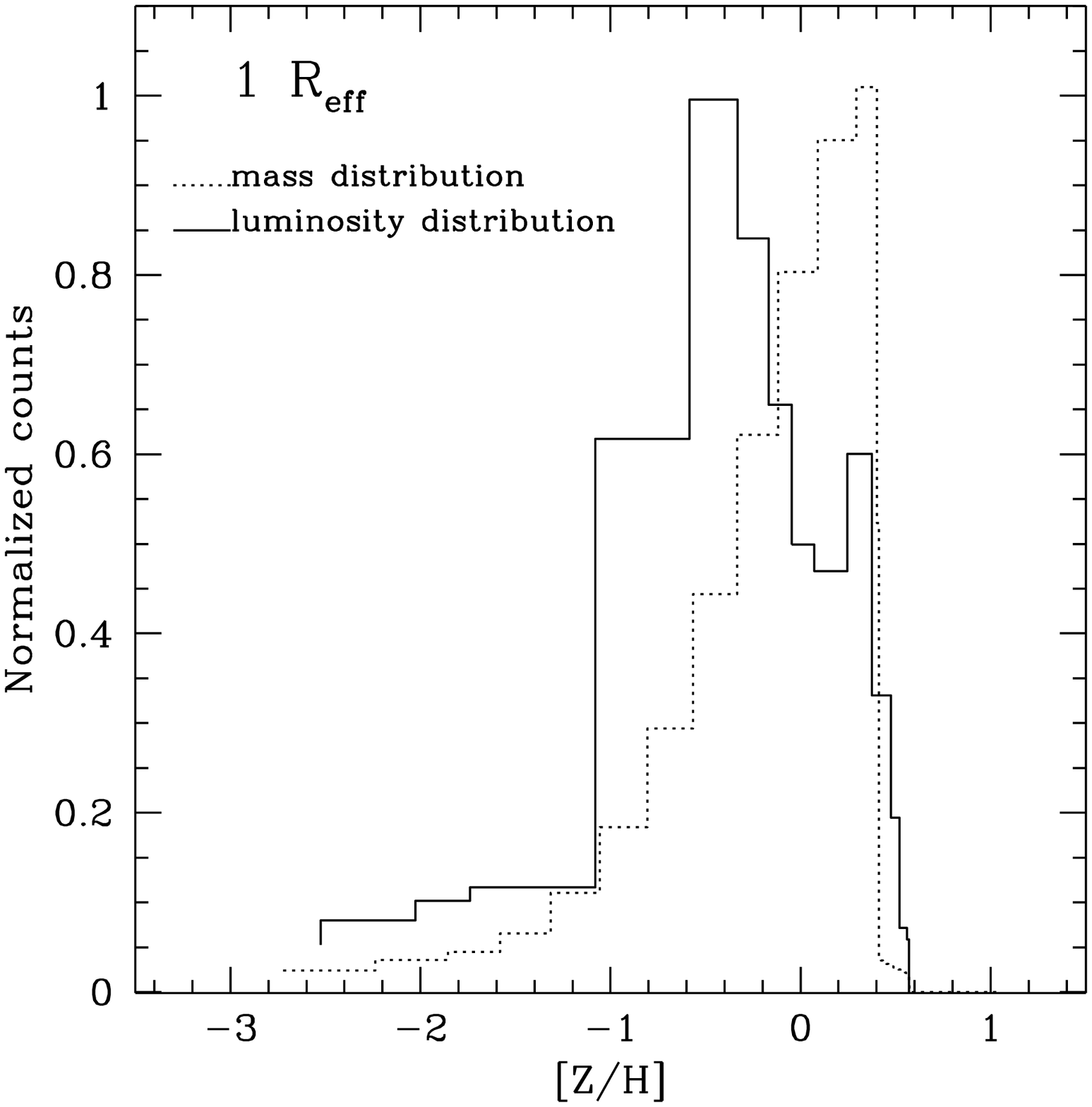}
\caption{``G-dwarf'' distributions for  
[Z/H] in luminosity (solid line) and mass (dotted line).
\emph{Left column}: values at $0.1 R_{eff}$.
\emph{Right column}: values at $1 R_{eff}$.
The plots are presented in the same scale in order to better
appreciate the differences among the different distributions.
}
 \label{Gdwarf}
\end{figure*}

From the comparison between our model predictions (Fig.~\ref{Gdwarf}) and the observed
G-dwarf-like diagrams derived at different radii 
by Harris \& Harris (2002, see their fig. 18) for the elliptical galaxy NGC 5128, we can derive some general considerations.
The qualitative agreement is remarkable: we can explain the slow rise in the [Z/H]-distribution
as the effect of the infall, whereas the sharp truncation at high metallicities is
the first direct evidence of a sudden and strong wind which stopped the star formation.
The suggested outside-in formation process reflects in a more asymmetric shape of the G-dwarf diagram
at larger radii, where the galactic wind occurs earlier (i.e. closer to the peak of the star formation rate),
with respect to the galactic centre.

\begin{figure*}
\centering
\includegraphics[width=3.5in,height=2.4in]{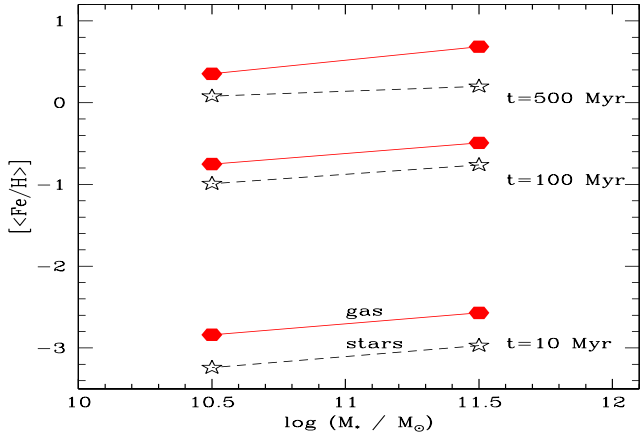}
\caption{The temporal evolution of mass-metallicity relation for the two studied galactic components (stars and gas).}
 \label{massmet}
\end{figure*}

From a quantitative point of view, properties suchs as the stellar metallicity distribution of the CSP inhabiting 
the galactic core, allow us to study
the creation of mass-metallicity relation (see Matteucci, this conference), which
is tipically inferred from the spectra taken at $\sim 0.1$ effective radius.
In Fig.~\ref{massmet} we plotted the time evolution of the mass-metalliticity relation
in the stars (which reflect the average chemical enrichment of the galactic core as seen at the present day; dashed line) and in the gas
(which, instead, is closer to the composition of the youngest SSP, thus being more indicative of a high redshift object; solid line).
The mean Fe abundance in the stellar component can reach the solar value in only 0.5 Gyr,
making ellipticals among the most metal-rich objects of the universe.

On the other hand, at variance with the G-dwarf-like diagrams as a function of [Z/H] (and [Fe/H]),
abundance ratios such as [$\alpha$/Fe] have  narrow and almost symmetric distributions. This means that, also
from a mathematical point of view, the [<$\alpha$/Fe>] ratio
are representative of the whole CSP (PMC06). The robustness of the [$\alpha$/Fe] ratios as constraints for
the galactic formation history is testified by the fact that 
$[\rm <\alpha/Fe>]\simeq [\rm <\alpha/Fe>_V]$, having very similar
distributions.
In particular, we find that the skewness parameter is much larger for the [Z/H] and [Fe/H]
distributions than for the case of the [$\alpha$/Fe] one, by
more than one order of magnitude. Moreover, the asymmetry
increases going to large radii (see Fig.~\ref{Gdwarf}, right panel), up to a factor
of $\sim$7 with respect to the inner regions. Therefore, it is not surprising that
the $[\rm <Z/H>]$ value does not
represent the galaxy at large radii,
and hence, we stress that care should be taken when
one wants to infer the real abundances of the stellar components
for a galaxy by comparing the observed indices (related
to a CSP) with the theoretical ones (predicted
for a SSP). Only the comparison based on the $[\rm <\alpha/Fe>]$ ratios
seems to be robust.


Another possible source of discrepancies is the fact
that luminosity-weighted averages (which are more closely related
to the observed indices) and mass-weighted averages (which
represent the real distributions of the chemical
elements in the stellar populations) might differ more in the most external
zones of the galaxy (compare the panels in Fig.~\ref{Gdwarf}).
All these considerations result in the fact that the chemical
abundance pattern used by modellers to build their SSPs, might
not necessary reflect the real trends.
Therefore, the interpretation of line strenght indices
in term of abundances, can be seriously flawed (see PMC06 for further details).

The analysis of the radial variation in the CSPs inhabiting elliptical galaxies seems to be promising as
a powerful tool to study ellipticals. Pipino, Puzia \& Matteucci (in preparation),
make use of the G-dwarf like distributions predicted by PMC06 to explain the multimodality in the globular
cluster (GC) metallicity distribution as well as their high $\alpha$ enhancement (Puzia et al. 2006). In particular, preliminary results
show that the GC distribution as function of [Fe/H] for the whole galaxy can be constructed
simply by combining distributions as those of Fig.~\ref{Gdwarf} (typical of different radii), once they had been rescale by means of a suitable
function (of time and metallicity) which links the global star formation rate
to the globular cluster creation. Neither a need of an enhanced GC formation during mergers
nor a strong role of the accretion of exteranl objects, seem to required
in order to explain the different features of the GC metallicity distributions.

Since globular clusters are the closest approximation of a SSP, we expect that this tecnique
will be very helpful to probe the properties of the stellar populations in spheroids,
thus avoiding the uncertainties typical of the analysis based on their integrated spectra.

\section{Concluding remarks}

A detailed study of the chemical properties of the CSPs inhabiting elliptical
galaxies as well as the change of their properties as a function of both time and radius, allow us to 
gather a wealth of information. Our main conclusions are:

\begin{itemize}
\item Both observed and predicted G-dwarf like distributions
for ellipticals show a sharp struncation at high metallicities that, in the light of our models, might
be interpreted as the first direct evidence for the occurrence of the galactic
wind in spheroids.
\item The stellar component of the most massive and luminous galaxies
might attain a metallicity $Z \ge Z_{\odot}$ in only 0.5 Gyr.
\item PM04's best model prediction of increasing [$\rm <\alpha/Fe>$] ratio
with radius is in very good agreement with the observed
gradient in [$\alpha$/Fe] of NGC 4697.
This strongly suggests an outside-in galaxy formation scenario
for elliptical galaxies that show strong gradients.
\item By comparing the radial trend of [$\rm <Z/H>$] with the 
\emph{observed} one,
we notice a discrepancy which is due to the fact that a
CSP behaves in a different way with respect to a SSP.
In particular the predicted gradient of [$\rm <Z/H>$] is flatter
than the observed one at large radii.
Therefore, this should be taken into account when 
estimates for the metallicity of a galaxy are derived
from the simple comparison between the observed line-strength index
and the predictions for a SSP, a method currently adopted in the literature.
\item Abundance ratios such as [Mg/Fe] are less affected by the discrepancy
between the SSPs and a CSP, since
their distribution functions are narrower and more symmetric.
Therefore, we stress the importance of such a ratio as the
most robust tool to estimate the duration of the galaxy
formation process.
\item Our results are strenghtened by the comparison between
our G-dwarf diagrams to metallicity distribution of the globular clusters residing in ellipticals.
\end{itemize}

The work was supported by MIUR under COFIN03 prot. 2003028039.
A.P. thanks the Organizers for having provided financial support for attending the conference.

\end{document}